\documentclass[final,5p,times,twocolumn]{elsarticle}

\usepackage{graphicx}
\usepackage{amssymb}
\usepackage{amsmath}
\usepackage{empheq} 
\usepackage{comment}
\usepackage{balance}
\usepackage{enumerate}
\usepackage[inline,shortlabels]{enumitem}
\usepackage[usenames,dvipsnames]{color}
\usepackage[colorlinks=true,linkcolor=Maroon,citecolor=Maroon,urlcolor=Maroon]{hyperref} 
\usepackage{url}

\biboptions{sort&compress} 

\makeatletter
\def\@author#1{\g@addto@macro\elsauthors{\normalsize%
    \def\baselinestretch{1}
    \upshape\authorsep#1\unskip\textsuperscript{%
      \ifx\@fnmark\@empty\else\unskip\sep\@fnmark\let\sep=,\fi
      \ifx\@corref\@empty\else\unskip\sep\@corref\let\sep=,\fi
      }%
    \def\authorsep{\unskip,\space}%
    \global\let\@fnmark\@empty
    \global\let\@corref\@empty  
    \global\let\sep\@empty}%
    \@eadauthor={#1}
}
\makeatother

\makeatletter
\def\ps@pprintTitle{%
 \let\@oddhead\@empty 
 \let\@evenhead\@empty
 \def\@oddfoot{}%
 \let\@evenfoot\@oddfoot}
 \makeatother
 
\newcommand{\nrm}[1]{\left| \left| #1 \right| \right|} 			
\newcommand{\M}{\mathbf{M}} 						
\newcommand{\A}{\mathbf{A}} 							
\newcommand{\FM}[1]{\mathbf{#1}} 						
\newcommand{\td}[2]{ \frac{\mathrm{d} #1}{\mathrm{d} #2}}  	
\newcommand{\dd}{\mathrm{d}} 						

\journal{~}
\begin{document}
\begin{frontmatter}

\title{Non-Normal Interactions Create Socio-Economic Bubbles}

\author[add1]{Didier Sornette}
\author[add1,add3]{Sandro Claudio Lera}
\author[add4,add5]{Jianhong Lin}
\author[add1]{Ke Wu}

\address[add1]{\scriptsize Institute of Risk Analysis, Prediction and Management (Risks-X), Academy for Advanced Interdisciplinary Sciences, Southern University of Science and Technology, Shenzhen, China }
\address[add3]{\scriptsize Connection Science, Massachusetts Institute of Technology, Cambridge, USA}

\address[add4]{\scriptsize Blockchain and Distributed Ledger Technologies, Institute of Informatics, University of Z\"urich, Andreasstrasse 15, 8050 Z\"urich, Switzerland}
\address[add5]{\scriptsize UZH Blockchain Center, University of Z\"urich, Andreasstrasse 15, 8050 Z\"urich, Switzerland}

\begin{abstract}
In social networks, bursts of activity often result from the imitative behavior between interacting agents. 
The Ising model, along with its variants in the social sciences, serves as a foundational framework to explain these phenomena through its critical properties. 
We propose an alternative generic mechanism for the emergence of collective exuberance within a broad class of agent-based models. 
We show that our model does not require the fine-tuning to a critical point, as is commonly done to explain bursts of activity using the Ising model and its variants. 
Instead, our approach hinges on the intrinsic non-symmetric and hierarchical organization of socio-economic networks. 
These non-normal networks exhibit transient and unsustainable surges in herd behavior across a wide range of control parameters even in the subcritical regime, thereby eliminating the need for the - arguably artificial - fine-tuning proximity to a critical point. 
To empirically validate our framework, we examine the behavior of meme stocks and establish a direct linkage between the size of financial bubbles and the degree of non-normality in the network, as quantified by the Kreiss constant.
Our proposed mechanism presents an alternative that is more general than prevailing conceptions of instabilities in diverse social systems.
\end{abstract}
\end{frontmatter}

\section*{Introduction}

Many complex dynamical systems are characterized by periods of relative stability which are 
interrupted by transient regimes during which the dynamics exhibits {sudden bursts of unsustainable growth} (a ``bubble'') or shifts suddenly to another attractor. 
A large corpus of knowledge and methods have been developed in the last two decades to account for these phenomena,
which are based on the underlying concept of tipping points, wherein a critical threshold is reached at which the system bifurcates to a new state 
\cite{Sornettebook2004,Cont2000,Schefferbook2009,Rocha2018,Bury2021d}. 
Typical financial models that account for such instabilities contain two classes of traders: 
fundamentalists who maximize their expected utility function  and noise traders \cite{DeLong1990,Scholl2021}. 
Noise traders are usually assumed to influence each other according to an Ising-like dynamics, 
with interaction dependencies captured by an adjacency matrix $\A$ and interaction strength captured by a coupling constant $\kappa$. 
When the imitation strength between noise traders is large enough, collective social behavior can occur, 
such as polarization of noise traders toward buying (selling), which in turn creates bubbles (crashes) \cite{Bouchaud2003,Sornette2017}.

These bubbles and crashes are generally associated with the underlying Ising phase transition separating 
a disordered opinion regime for low imitation strength $\kappa$ from an ordered regime where all noise traders tend to be synchronized.
In all existing models of this type, 
bubbles requires the imitation strength $\kappa$ to be close to or larger than a critical value $\kappa_c$ associated with the underlying phase transition. 
In other words, in this class of models, bubbles and crashes are the signatures of the fact that the financial market has entered  a ``critical regime'', 
in the technical sense of the emergence of collective order in the decisions of a large fraction of traders.
There is an extensive literature on agent-based models and generalized Ising models 
\cite{Mantegna1999,Galam2008,Buldyrev2010,Scheffer2012,Battiston2016}.
To the best of our knowledge - in all cases - 
the abnormal stylized facts, such as excess volatility and transient bubbles and crashes, require the system to be close to, at or slightly above the critical point in the ordered polarizing regime.

In this work, we document a general mechanism for the nucleation and growth of transient bubbles.
We suggest that this new mechanism is much more general than previously existing ones and likely to be often the dominant process at work, 
because it does not require the fine-tuning close to or sweeping \cite{SornSweep94} of the system over a critical point. 
Our proposed mechanism is based on the fact that social influence is characterized by two critical properties: it is directed and hierarchical \cite{Corominas2013,LeraSorn19}.
For instance, in our example of financial markets, the influence of a famous investor on a retail investor is likely much larger than the other way around. 
Together, these two ingredients give rise to networks with non-normal adjacency matrices $\A$, 
called non-normal networks \cite{Asllani2018b,OBrien2021,Duan2022}, 
whose associated dynamical systems are known to induce transient bursts 
\cite{Trefethen1993,Embree2005,Murphy2009,Biancalani2017}.
Interpreted in terms of socio-economic interactions, these transient bursts are responsible for short-lived social contagion even well below any critical threshold.
We demonstrate this mechanism in the context of the formation of financial bubbles and their following crashes that lead to enormous economic losses. 
Analyzing Reddit discussion forums of meme stocks, we show that patterns of influence are highly non-normal, and that the rate of reciprocity is dependent on a user hierarchy. 
Using a previously validated agent-based model \cite{kaizoji2015super,westphal2020market},
we show that non-normal networks give rise to transient bubbles even when the imitation strength is sub-critical. 
Intuitively, some traders are more influential than others and information does not spread evenly but along cascading circuits.
Our work thus provides a qualitative proposal that financial systems are intrinsically generating crises \cite{Minsky1957}.
Due to the broad applicability of stochastic interaction models \cite{Castellano2009}, 
our model is expected to generalize in a straightforward manner to other hierarchical socio-economic systems,
and help explain wide-spread phenomena such as social bubbles \cite{GislerSor11} and herding in opinion dynamics \cite{Galam2008}.

\section*{Results and Discussion}

\subsection*{Agent Based Price Simulation}

Agent-based models (ABMs) have become a popular tool in interdisciplinary research over the last decades \cite{Sornette2014,Dosi2019,Ott2022}, 
primarily due to their flexibility in accounting for heterogeneous and non-linear interactions. 
Here, we implement an ABM that simulates a financial market consisting of fundamentalists and noise traders who trade a risky and a risk-free asset \cite{kaizoji2015super,westphal2020market}.
The risky asset is a dividend paying stock. 
The risk-free asset pays a constant return in each time-step and represents a bank account or risk-free government bond. 
Each trader formulates their excess demand for the next time step and the price $P_t$ of the risky asset is calculated as the Walras equilibrium in which supply equals demand. 
Fundamentalists are rational risk-averse investors who invest by maximizing their expected utility under a constant relative risk aversion utility function.
At each time step, they allocate their wealth between the two assets in order to maximize their expected utility over the next period.
They buy (sell) or sell (buy) the risky (riskless) asset in order to move their portfolio towards their desired
optimal portfolio, given the information they have on price and volatility of the risky asset.
They play a role akin to price-setters, in contrast with noise traders who are pure price-takers.
Their demand for the risky asset combines with that of the noise traders to fix the price at each time step.
We refer to the Supplementary Notes for more details on the fundamentalists. 
\footnote{
The Supplementary Notes are available from the authors upon request.
}

The crucial component for our model are the noise traders. 
Their investment strategy is based on an Ising-like social influence model, where they can be modeled as nodes in a network with directed edges.
While previous implementations \cite{kaizoji2015super,westphal2020market} have considered a mean-field approximation in terms of a fully connected, symmetric network, 
here we make the network topology an explicit component of our model. 
As will become clear below, relaxing the assumption of fully connected, symmetric interactions is not only realistic, but also a crucial ingredient for the observation of sub-critical bubbles.
Each trader $i$ is considered to be in one of the two possible states, 
$+1$ (the noise trader holds the risky asset), 
and $-1$ (the noise trader holds the risk-free asset). 
The states are denoted as $s^{i}=\pm{1}$, respectively. 
The directed network of $N$ nodes (noise traders) is described by its adjacency matrix $\A= \{ a_{ij} \}$, where $a_{ij}=1$ if there exists a directed edge (influence) from node $j$ to node $i$, and $a_{ij}=0$ otherwise. 
Introducing further the transition matrix $\M$ which is proportional to $\A$ (see Methods section for details)
allows us to represent the state dynamics of the Ising-like noise traders as 
\begin{equation}
    \label{eq:vectorial_state_function}
    \triangle{\vec{s}(t)}= \M \vec{s}(t)
\end{equation}
At each time step $t$, the collective opinion (``magnetization'' in the Ising language) $m_t$ is defined as
\begin{equation}
\label{eq:magnetization}
    m_t = \frac{1}{N} \sum_{i=1}^N s^i_{t}~~~\in[-1, 1].
\end{equation}
This system remains stable as long as the real parts of all eigenvalues of $\M$ are negative.
As is well known, by continuously tuning $\M$, 
systems whose linear stability is controlled by \eqref{eq:vectorial_state_function} can undergo a bifurcation (or phase transition) from a stable fixed point with zero average change of spin to a state where all spin change state to align to each other 
(a state described by higher-order terms beyond the linear expansion $\M\vec{s}(t)$).
The existence of such states has been related to the emergence of financial bubbles (crashes), diagnosed by the existence of transient super-exponential growth (loss) \cite{sornette2014financial,Sornette2017}. 
System \eqref{eq:vectorial_state_function} is well-known to be a generic representation of the aggregate behavioral outcomes of financial markets where investors are driven by group psychology and herding \cite{DeLong1990,Brock1998,Brock2001,Chiarella2009}.
Moreover, as will become clear below, our results are not restricted to the detailed assumptions of our model and generalize to a large class of stochastic interaction models (from hereon referred to as `Ising-like').
In the remainder of this article, we will analyze the sub-critical regime of $\M$, but with $\M$ being non-normal. 
We will show that this non-normal structure gives rise to transient dynamics that induce bubbles and crashes much like at or close to criticality in normal networks.

\begin{figure*}[!htb]
	\centering	\includegraphics[width=1\textwidth]{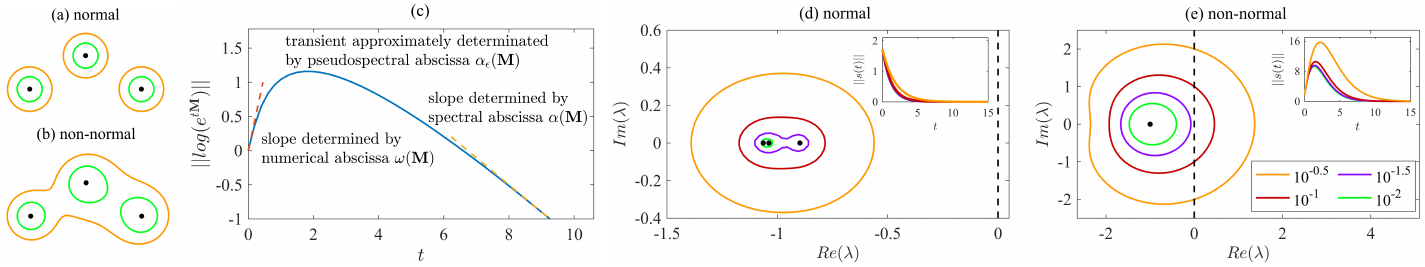}
	\caption{	\textbf{Difference between normal and non-normal transients.} 
			(a) and (b) show the geometry of pseudo spectra. 
			In each plot, the contours represent the boundary of $\sigma_\epsilon(\M)$ for two values of $\epsilon$ (see Methods).
			(c)
			Initial, transient and asymptotic behavior of $\nrm{e^{t \M}}$ for a non-normal matrix $\M$. 
			 The graph indicates that here $\sigma(\M) < 0$ and hence that the asymptotic behavior is stable. 
			 However, the asymptotic behavior is not at all predictive of the transient behavior in case $\M$ is non-normal.
			Plot (d) shows the eigenvalues (black dots) and some $\epsilon$-pseudospectra for a normal matrix
			(different colors represent different values of $\epsilon$). 
			All eigenvalues as well as the epsilon-spectral lines are confined to the left half plane of $\mathbb{C}$. 
			Accordingly, $||s_t||$ decays exponentially as shown in the inset plot. 
			Plot (e) shows the case of a non-normal matrix $\M$. 
			While its eigenvalues are also confined to the left half plane of $\mathbb{C}$, its $\epsilon$ spectral lines are not. 
			According to inequality \eqref{eq:Kreiss_inequality}, in the inset plot, we see intermittent transient growth before the asymptotic decay
			as in Plot (c). 
			}
	\label{fig:non_normal_theory}
\end{figure*}

\subsection*{Non-Normal Matrices}
\label{sec:primer_non_normality}

Early contributions to the study of non-normal matrices have originated from hydrodynamics, where non-normality plays a role in the emergence of turbulence \cite{Trefethen1993}.
Ever since, non-normal matrix theory has helped explain phenomena such as 
perturbations in ecosystems \cite{Neubert1997}, 
non-Hermitian quantum mechanics \cite{Hatano1996},
population dynamics \cite{Neubert2002}, 
synchronization of optoelectronic oscillators \cite{Ravoori2011}, 
amplification of neural activities \cite{Murphy2009}, 
chemical reactions \cite{Nicoletti2019}, 
network synchronization \cite{Asllani2018b,Nicoletti2018,Muolo2020}, 
neuronal networks \cite{Hennequin2012,Gudowska2020}, 
ecological reactions \cite{Tang2014}, 
and the formation of Turing patterns \cite{Biancalani2017,Muolo2019}. 

A matrix $\M$ is called \textit{non-normal} if $\M^T \M \neq  \M  \M^T$. 
Non-normal matrices, unlike their normal counterparts, contain at least one set of eigenvectors which are non-orthogonal to each other.
We refer to the Methods section for a primer on non-normal matrices. 
For the purposes at hand,
one key property of non-normal matrices $\M$ is that the dynamics they describe via \eqref{eq:vectorial_state_function} is significantly different from the long-term asymptotic behavior governed by the largest eigenvalue. 
It is well-known that the asymptotic behavior for $t \to \infty$ is governed by the largest real-part of all eigenvalues of $\M$. 
In particular, so long as all real parts of $\M$ are negative, the dynamics is stable. 
Our main interest lies in transients, which are described by intermediate values of $t$.

It can be shown that, for intermediate times, there is transient growth according to 
\begin{equation}
	\sup_{t \geqslant 0} \nrm{ e^{t \M} } \geqslant \mathcal{K}(\M). 
	\label{eq:Kreiss_inequality}
\end{equation}
Importantly for our application below, this initial transient growth is exponential (Figure \ref{fig:non_normal_theory}(c)).
Given an interaction matrix $\M$ as in \eqref{eq:vectorial_state_function}, 
we can calculate the Kreiss constant $\mathcal{K}(\M)$ (see Methods) to obtain lower bounds for the transient growth of net magnetization. 
An example of such transient growth is shown in Figure \ref{fig:non_normal_theory}(c) and in the inset of Plot (e). 
As we shall see, these transients are responsible for socio-economic bubbles in our agent-based model.

\subsection*{Level-Dependent Reciprocal Connections}
\label{sec:parametrization}

A system such as eq. \eqref{eq:vectorial_state_function} with non-normal $\M$
can be interpreted as a dynamical process on a complex asymmetric and partly hierarchical network.
Such non-normal networks have been observed in a wide variety of biological and socio-economic networks \cite{Asllani2018,OBrien2021,Duan2022}, 
and their role in the transmission of noise has been studied  \cite{Baggio2020}. 
Recall that asymmetry of $\M$ is a necessary, but not a sufficient condition for non-normality. 
For instance, consider a simple cyclical, directed network of three nodes $\{X,Y,Z\}$ where $X \to Y, Y \to Z$ and $Z \to X$. 
An adjacency matrix with such cyclical symmetry still gives rise to a normal adjacency matrix. 
The condition $\M \M^T \neq \M^T \M$ is instead satisfied when the network is directed and hierarchical, 
which are both intrinsic properties of socio-economic systems \cite{LeraSorn19,Johnson2020,Kawakatsu2021}. 
Indeed, it is well-established in anthropology that people organize their social interactions along one or more
of the four elementary modes: 
communal sharing, authority ranking, equality matching, and market pricing, with the latest one being particularly relevant in finance \cite{Fiske1992,MarouDid}.
Of these four modes, only equity matching is symmetrically directed, while the others exhibit some direction
asymmetry and a degree of hierarchy.
Assuming that non-normal matrices govern the interaction of humans in socio-economic and financial systems is thus natural.

Recently, methods to generate non-normal networks have been proposed by taking into consideration asymmetrical reciprocity \cite{Asllani2018} and hierarchy \cite{OBrien2021} that are typical of non-normal systems. 
Drawing from these insights, we implement here an algorithm that allows us to control the rate of non-normality along with the number of top nodes, that can be interpreted as thought leaders.
In contrast to previous work \cite{Asllani2018}, our rate of reciprocity explicitly depends on the hierarchical level which is a realistic addition as has been evidenced in a variety of biological and social non-normal networks \cite{OBrien2021}. 

\begin{figure*}[!htb]
    \centering
    \includegraphics[width=\textwidth]{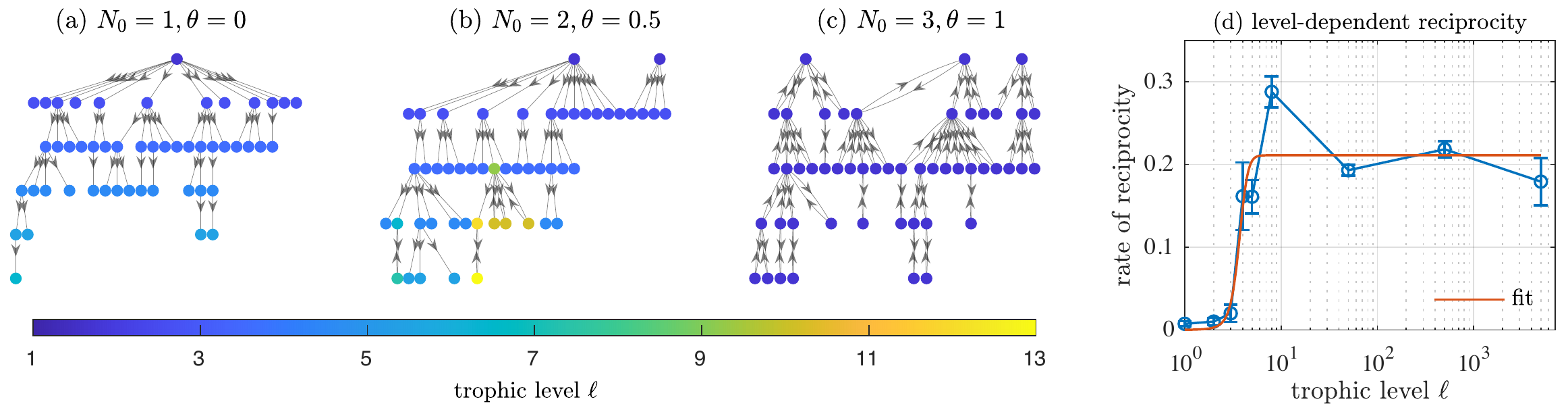}
    \caption{
    		\textbf{Hierarchical levels and rates of reciprocity in complex networks.}
    		(a-c)
    		Three examples of directed networks with different levels of non-normality, different number of top nodes, and different {but constant rates of reciprocity $\rho$ given by different values of $\theta$}. 
		The colors indicate the hierarchical (in general non-integer) level $\ell$ of the nodes. 
		A value of $\theta=0$ means no edge can be reciprocated, whereas a value of $\theta=1$ means every edge is reciprocated (since level-dependence is ignored). 
		Network (c) is then normal, since it is symmetric. 
		(d)
		Empirical analysis of the Blackberry subreddit network. 
		The rate of reciprocity is not constant, but a function of the hierarchical level, {$\rho = \rho(\ell)$}.
		The higher the level, the higher the rate of reciprocity, up to some level of saturation. 
		Error bars represent standard deviations of average rate of reciprocity measured across all users. 
		The red line shows the sigmoid function that best fits the data. 
            }
    \label{fig:example_network}
\end{figure*}

The non-normal network with a total of $N$ nodes is initialized with $N_0$ so-called top nodes. 
These top nodes account for the largely independent $N_0$ backbones of the communication network common to typical hierarchically, non-normal networks \cite{Asllani2018,OBrien2021}. 
The remaining $N-N_0$ nodes are added to the existing network sequentially, one node at a time. 
Each newly added node receives $m$ in-edges, i.e. channels of communication through which it can be influenced. 
The source of each such edge is selected with probability proportional to the existing nodes' out-degree.
As is well-known, this type of preferential attachment creates a skewed degree distribution whereby the network is dominated by a few central nodes \cite{Barabasi1999,Lera2020}. 
Once the $m$ source nodes of an added node are determined,  each of the $m$ newly formed directed edges may be reciprocated with some probability $\rho$. 
Figure \ref{fig:example_network}(a-c) exemplifies several cases where $\rho$ is equal to a fixed value $\theta$. 
Setting $\theta \ll 1$ gives rise to strongly non-normal systems \cite{Asllani2018}, whereas $\theta = 1$ recovers a symmetric, i.e. normal, network.

Based on empirical evidence \cite{OBrien2021}  and to reflect the fact that nodes that are higher up in the hierarchy are harder to be influenced, 
we assume further that this probability $\rho$ is modulated by the hierarchical level $\ell$ of each node, $\rho = \rho(\ell)$.
The lower the node in the hierarchy (the larger $\ell$), the more likely the node is reciprocated. 
Loosely speaking, the hierarchical level $\ell$ of any node $i$ is defined as the shortest path from a top-node to node $i$. 
More precisely, the level $\ell$ is defined as the trophic hierarchical level \cite{Johnson2017,Pilgrim2020}. 
We have analyzed the Reddit discussion forum for the Blackberry meme stock below. 
An edge is drawn from user $i$ to user $j$ if $j$ replies to a comment of user $i$. 
In Figure \ref{fig:example_network}(d), we show that the rate of reciprocity is not constant, but an increasing function of $\ell$. 
In other words, the more popular a user's comments, the less likely that user is to reciprocate (comment on) any given edge.
For social communication networks, this observation is natural and rationalized as the approximately constant finite capacity of any given individual to respond to comments. 

We detail in the Methods section our algorithm to generate non-normal adjacency matrices.
Such a matrix $\A$ has six parameters: $N, N_0, m, \theta, a$ and $b$. 
The four parameters $m, N, a$ and $b$ play a subordinate role in the qualitative interpretation of our results. 
For the remainder of this paper, we thus fix $N=1000$, $m=2$, $a=2.552$ and $b = 3.668$ unless mentioned otherwise. 
The parameters $\theta$ and $N_0$, on the other hand, have qualitatively important implications on the behavior of our model. 
The parameter $\theta$ characterizes the asymmetric nature of the system. 
The smaller $\theta$, the more directed the network, and the less the top nodes may be influenced.
The parameter $N_0$ denotes the number of top nodes. 
If $\theta$ is small, then $N_0$ may be interpreted as the number of (largely) independent, leading opinions in the system.
As we will now show, our results do not depend on the specific implementation of the non-normal mechanism, but are primarily mediated by the Kreiss constant
(see Primer on Non-Normality in the Methods section).

\subsection*{Transient Bubbles Induced by Non-Normal Interactions}
\label{sec:non_normal_interactions}

\begin{figure*}[t]
\centering\includegraphics[width=\textwidth]{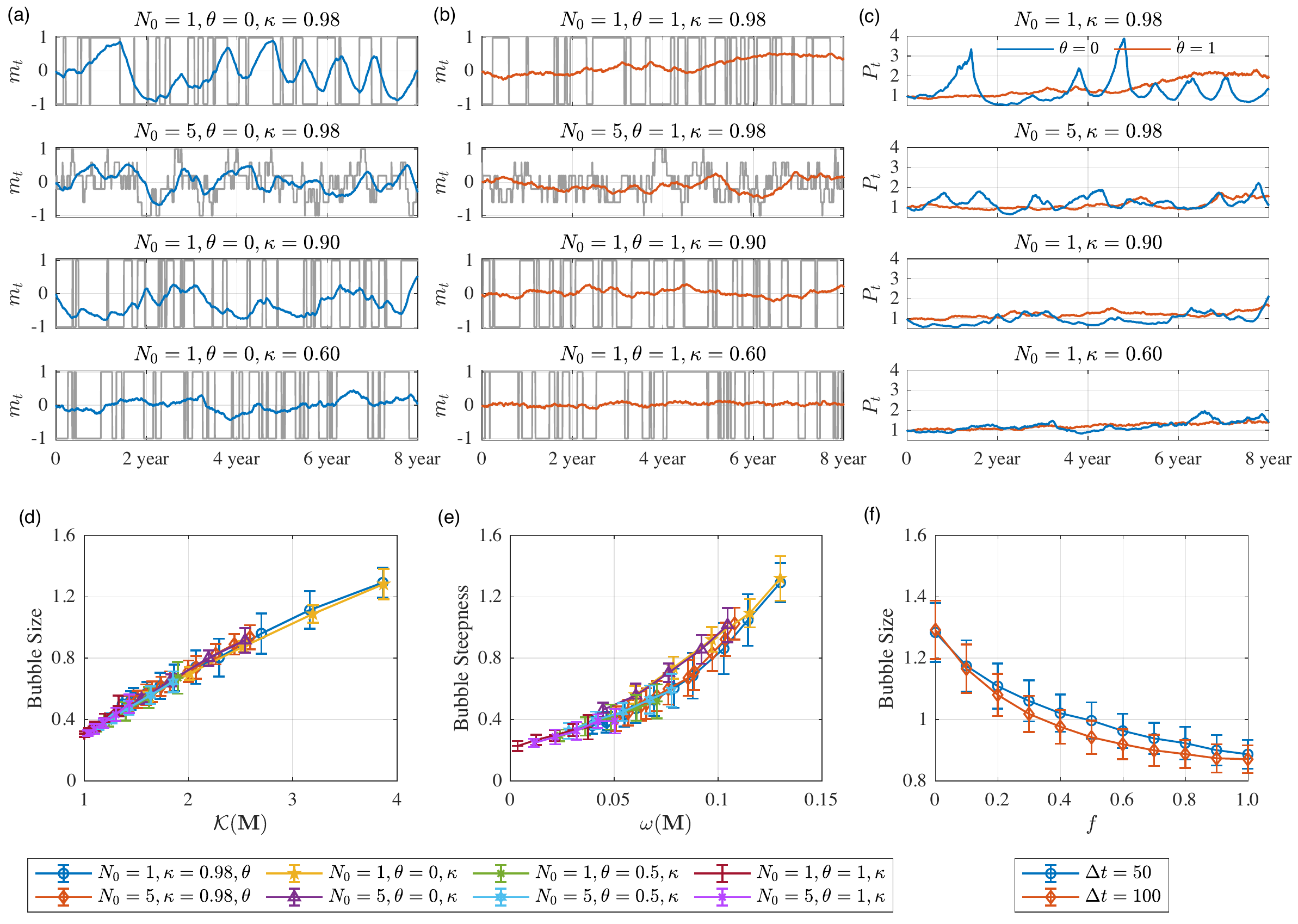}
\caption{
\textbf{Normal vs. non-normal price dynamics.}
(a)
Magnetization \eqref{eq:magnetization} for an ABM simulation with a non-normal interaction matrix $\M$ for different parameter constellations. 
The net magnetization of the $N_0$ opinion leaders is shown in grey lines in the background.
The opinion leaders change their opinion randomly at rate $p^\pm / 2$, as dictated by equation \eqref{eq:trans_probability}.
(b) 
Same as in (a), but for a symmetrized (i.e. normal) interaction matrix $\frac{1}{2} \left( \M + \M^T \right)$. 
The parameter $\kappa$ is chosen at a sub-critical value, hence the net magnetization is, on average, $0$. 
In contrast to (a), the transient bursts are much less pronounced. 
(c)
Price trajectory generated by the agent-based model dynamics with magnetization from (a) and (b), respectively. 
Only the non-normal matrices induce bubbles.
(d)
Bubble size as a function of Kreiss constant for different parameter constellations. 
(e)
Bubble steepness as a function of numerical abscissa for different parameter constellations. 
(f)
Bubble size as a function of fraction of nodes receptive to antagonistic opinion. 
Error bars represent standard deviations obtained from 100 simulations.
}
\label{fig:ABM_bubbles}
\end{figure*}

Building on the above insights, we now run agent-based simulations with a non-normal adjacency matrix $\M$. 
While it has been well-established that the formation of bubbles in agent-based models with Ising-like interactions
is associated with the proximity to the critical point of the underlying Ising model with the occurrence
non-zero net opinion (non-zero magnetization) \cite{Lux1995,jiang2010bubble,Sornette2017}, 
we investigate here the regime where the net magnetization (net opinion) fluctuates around zero (sub-critical phase).
In the following analysis, we therefore set the coupling strength $\kappa$ to a sub-critical value $(\kappa=0.98)$. 
The eight panels of Figure \ref{fig:ABM_bubbles}(a,b)  confirms that $m_t$ fluctuates around zero approximately symmetrically, 
as expected from the fact that the imitation strength $\kappa$ has been chosen so that the underlying Ising model is subcritical.
Furthermore, for fixed parameters  $(N_0, \kappa)$, we compare two types of social networks: $\theta = 0$ and $\theta=1$. 
The former corresponds to a case of no reciprocity, and hence large non-normality of $\A$ and hence $\M$. 
The later corresponds to an almost symmetric - and hence normal - matrix $\A$, which coincides with a much less non-normal matrix $\M$ 
(see also Supplementary Notes).
Comparing Figure \ref{fig:ABM_bubbles} (a) and (b), 
we see that the strongly non-normal case (small $\theta$) corresponds to much more pronounced long-lived deviations of the magnetization from its zero average, 
as is expected from transient dynamics (Figure \ref{fig:non_normal_theory}(c) and (e)). 

The right most column of Figure \ref{fig:ABM_bubbles} shows 
the associated price $P_t$ as a function of time, obtained as the Walras' equilibrium between fundamentalists and noise traders (Supplementary Note 1). 
It is striking to observe the drastic differences in the price dynamics between  highly non-reciprocal (non-normal) interactions compared to reciprocated (normal) ones. 
In the former, very strong price peaks are preceded by periods of strong price growth, following by fast large asymmetric drawdowns.
This qualifies the existence of large amplitude bubbles as a clear diagnostic of this type of non-normal networks. 
In contrast, for normal networks (and weakly non-normal $\M$ matrices),  the price dynamics appears compatible with a standard geometric Brownian motion at least at the qualitative level.
As we now show, both of these phenomena are explained as a function of the transients induced by non-normality.

A hallmark of a financial bubble is the existence of unsustainable super-exponential price growth \citep{Sornette2017,sornette2014financial,Husler2013}.
{The term super-exponential refers to a special regime where the price grows much faster than exponential, with 
its growth rate growing itself as a function of time, compared with exponential growth characterized by a constant growth rate.}
Within our ABM, it can be shown \cite{kaizoji2015super} that, to a first approximation, the price is an exponential function of the net magnetization, 
$
 P_t = C e^{c m_t}, 
$
where the scaling coefficient $c>0$ is a function of the model parameters. 
This result is a direct consequence of the fact that, on short time-scales, the price is mostly driven 
by the demand of noise-traders.
Thus, our results are not specific to the details of our implementation of the social imitation model and are expected to hold for general classes of stochastic interaction systems \cite{Liggett1999} and common assumptions about fundamentalists \cite{Brock2001}.

We recall that $m_t$ is defined as the average state across all trader states $\vec{s}(t)$. 
The states $\vec{s}(t)$ are themselves governed by equation \eqref{eq:vectorial_state_function} involving the non-normal interaction matrix $\M$, such that $\vec{s}(t) \sim e^{\M t} \vec{s}(0)$. 
In a globally stable regime, all eigenvalues of $\M$ associated with the stable equilibrium $\vec{s}(t) = 0$  are negative (Supplementary Note 1).
The standard expectation is thus that $m_t$ remains small and thus $P_t$ should not exhibit abnormal fluctuations.
But this is forgetting the transients induced by the non-normality of $\M$.
As shown in Figure \ref{fig:non_normal_theory}(c,e), 
the asymptotically stable fixed-point $\vec{s} = 0$ is punctuated by {initially exponential}, repelling dynamics over finite time scales. 
Furthermore, inequality \eqref{eq:Kreiss_inequality} provides us with a lower bound of the size of these transients, which are mainly a function of the Kreiss constant $\mathcal{K}(\M)$. 
Combining $P_t = C e^{c m_t}$ with the transient approximately exponential growth of $m_t$,  we thus predict the occurrence of finite lived bubbles qualified as transient super-exponential growth of price. Here, the super-exponential behavior is approximately described by an exponential of an exponential.

This reasoning leads us to hypothesize that various properties of emergent bubbles should be mostly driven by metrics characterizing the non-normal nature of the network of interacting noise traders. 
The most obvious candidate to characterize non-normal networks being the Kreiss constant $\mathcal{K}(\M)$,
we hypothesize that the dependence of the size of emerging bubbles on parameters $(N_0, \theta, \kappa)$ should reduce to a sole 
dependence on $\mathcal{K}(\M)$. 
To test this dependence, we measure the size of the bubbles as the difference in price between the beginning and the end of a regime of super-exponential growth  (see Supplementary Note 5 for details). 
It turns out that the sizes of bubbles are almost entirely explained by just the Kreiss constant, for various parameter constellations, as shown in Figure \ref{fig:ABM_bubbles}.

More precisely, for different parameter combinations of  $(N_0, \theta, \kappa)$, we generate $100$ price simulations according to the following procedure. 
We first generate a matrix $\M$ according to our algorithm described above, 
and we subsequently run the agent-based model to generate 
a time-series with $25,000$ time-steps, corresponding to $100$ years (considering 250 trading days per calendar year). 
On each time series, we measure the size of all bubbles. 
These sizes are subsequently averaged across all 100 simulations, with the standard deviation serving as error bars. 
Figure \ref{fig:ABM_bubbles}(d) demonstrates the existence of a collapse of all curves when the average bubble sizes are plotted as a function of the Kreiss constant $\mathcal{K}(\M)$ of the non-normal matrix $\M$.
A large Kreiss constant is associated with large bubble sizes, for different network non-normality and social coupling $\kappa$.

The theory of transients does not only make a prediction about the size of the transients, but also about their steepness.
As visualized in Figure \ref{fig:non_normal_theory}(c), we expect the steepness of the transients, and therefore of the magnetization and then price, to be increasing in the numerical abscissa $\omega(\M)$. 
Figure \ref{fig:ABM_bubbles}(e) shows that the bubble steepness is indeed mostly explained by the numerical abscissa $\omega(\M)$ of the non-normal network, for different sets of parameters. 

These results underline the crucial role of the non-normality of the noise traders interaction network in shaping the bubbles and their properties in our agent-based model.
The key insight is that different parametrization indeed all collapse onto this scaling law relating the bubble size to the Kreiss constant
and the bubble steepness to the numerical abscissa of the matrix $\M$ characterising the social network of noise traders.
The intuition behind this mechanism is related to the hierarchical organization associated with non-normal matrices \cite{OBrien2021}. 
The information flow is structured such that it is driven by a few leaders, thereby reducing the effective number of independent opinions and increasing the level of polarization.

Finally, we test the effect of an influential contrary opinion on the size of the bubble. 
We first simulate a price dynamics with a single top node, $N_0=1$, with $N=1000$, $m=2$ and $\theta=0$. 
Upon formation of a bubble (continuous price increase for 50 time-steps), 
we interfere with the system by holding a contrary opinion (opposite state of $N_0$) for $\Delta t$ time-steps. 
That contrary node is connected to a fraction $f$ of all nodes with the exception of the top node. 
Figure \ref{fig:ABM_bubbles}(f) shows the dependence of the bubble size as a function of $f$. 
The larger $f$ is, the larger is the decrease of the bubble size. 
This result is encouraging, suggesting that an external controller can partially counteract the development of sub-critical bubbles.
However, this control remains limited even when $f \to 1$ and requires one to reach to a large fraction of noise traders.
A more scalable approach, that we leave for future research, would be a minimal influence of a few key nodes in order to achieve overall noise-cancellation, as has been recently shown in communication networks \cite{Baggio2020}, 
and which could improve on the more standard market intervention involving large balance sheet build-up of major financial agents such as a central bank \cite{Miao2018}.

\subsection*{Non-Normal Communication in Meme Stock Trading}
\label{sec:MemeStocks}

\begin{figure}[!htb]
	\centering
	\includegraphics[width=0.47\textwidth]{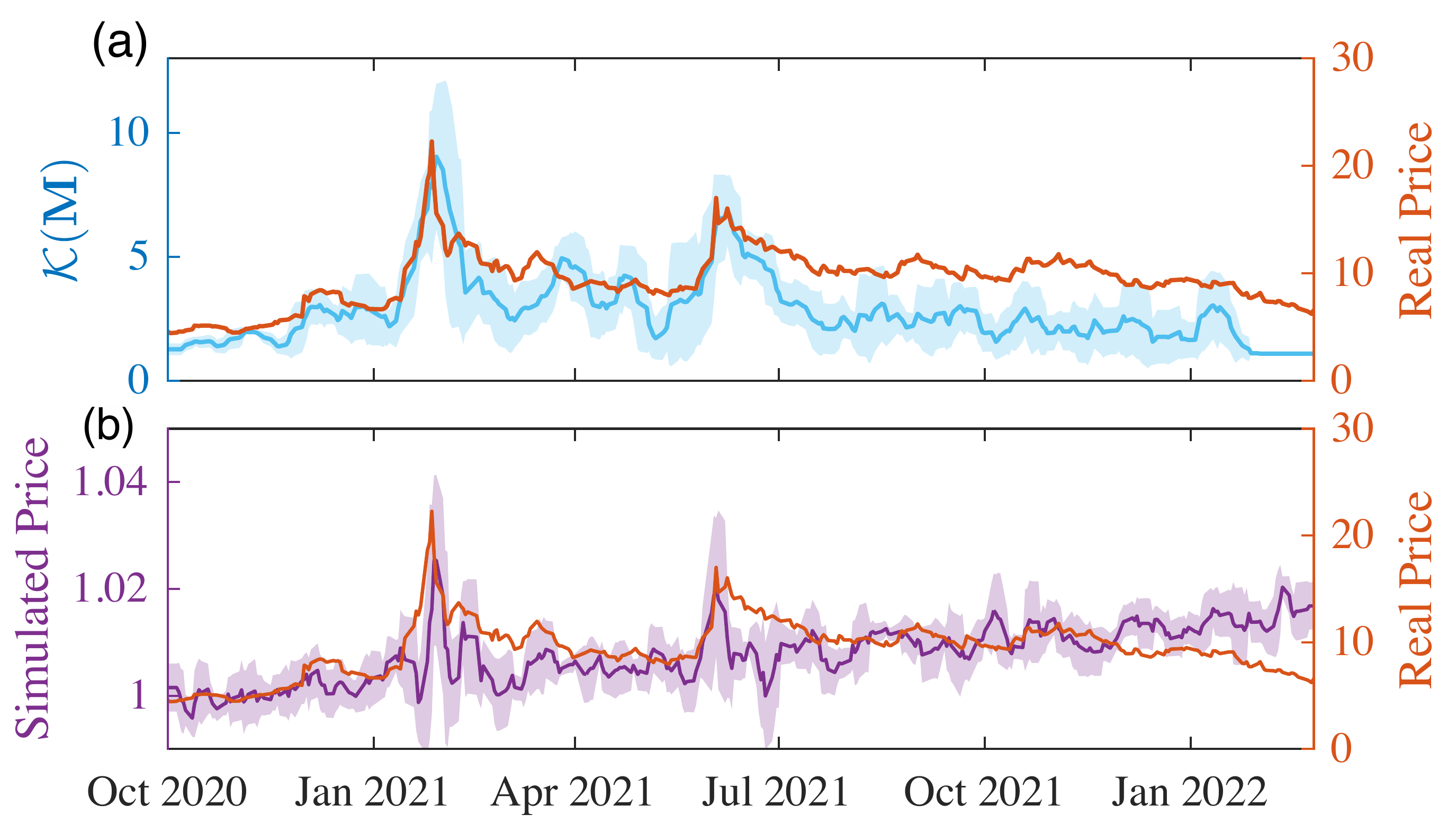}
	\caption{
	\textbf{Evolution of real and simulated meme stock prices based on discussions on Reddit forums forming non-normal networks of influence.}
	(a)	Co-evolution of Reddit network Kreiss constant and Blackberry meme stock price. 
	(b)	ABM price time-series resulting from the simulated magnetization $m_t$ of noise traders, with distinct peak around the same time as the real price trajectory.
	The bands around the thick lines represent confidence intervals obtained across 100 simulations.
	}
	\label{fig:empirical}
\end{figure}

So far, our assessment of bubbles has relied on agent-based simulations, where we can control the experimental conditions.  
The difficulty with empirical data is that, in general, one cannot observe the matrix $\A$ that governs trader interactions. 
On social trading platforms, such as eToro, interactions can be measured precisely \cite{Lera2020}, 
but the trading volume relative to the entire market is small, such that its influence on the price is negligible. 
This is not the case for so-called meme stocks which have enjoyed recent popularity. 
Driven primarily by retail traders, meme stock trading activity has been shown to be largely influenced by Reddit discussion forums \cite{Lyocsa2021,Lucchini2022,Betzer2022}. 

Reddit is organized into \textit{subreddits} on which specific topics are being discussed. 
Users interact by submitting new posts and adding comments to existing posts or comments. 
Here, we analyze the posts and comments related to four popular meme stocks (Blackberry, Nokia, GameStop and AMC) under the famous subreddit \textit{r/wallstreetbets} (also known as WallStreetBets or WSB) that has become notable for its colorful and profane jargon, aggressive trading strategies, and for playing a major role in the GameStop short squeeze in early 2021. For each of the four stock, at time $t$, we draw a directed edge from user $J$ to user $K$ if $K$ has commented or replied a stock-related text by user $J$ in the time interval $[ t-\Delta t, t]$. 
In other words, $K$ has been influenced by $J$'s action in the past $\Delta t$ days. 
With this procedure, for each meme stock, we extract a dynamically evolving influence network $\A(t)$, for which we can measure the Kreiss constant $\mathcal{K}(t)$. 

The evolution of the Kreiss constant, along with the trading price, is shown in the top plot of Figure \ref{fig:empirical} for the Blackberry stock (see Supplementary Note 6 for similar plots for the other three meme stocks). 
The two most prominent price peaks around January 2021 and June 2021 coincide with the two largest peaks of the Kreiss constant trajectory.
This gives force to our proposition that increased non-normality (quantified by large values of the Kreiss constant)
favours the occurrence of transient {super-exponential} price behavior (bubbles) associated with the transient growth of perturbations before their relaxation. 
In other words, we interpret the presence of financial bubbles in these meme stocks as reflecting at least partially the asymmetric hierarchical structure of the reddit discussion forum that induced a polarized bullish opinion among retail traders, 
which then in-turn pushed the price up. 
And indeed, the mostly mentioned words in Jan 2021 among the reddit submissions and comments related to BlackBerry are 
\textit{rocket}, 
\textit{bb}, 
\textit{gme}, 
\textit{shares}, 
and \textit{buy}.
In particular, among the 58,793 mentions of ``rocket'' and 12,132 mentions of ``buy'' in 2021, 69\% of ``rocket'' and 43\% of  ``buy'' were in January. 
More generally, as is shown in Supplementary Note 6, we do find a positive correlation between the Kreiss constant and price bubbles across meme stocks. 
This suggests that the non-normal structure of the Reddit meme stock discussion forum is an important driver of the observed price instabilities. 

A strong asymmetric hierarchical structure of the reddit discussion forum quantified by a large value of the Kreiss constant provides a powerful catalysis for the emergence of transient price bubbles. 
But in a world of mass communication and a plurality of social media and information channels, not all observed price perturbations can realistically be expected to be attributed to the discussions on Reddit forums. 
As for general dynamics with non-normal operators, not all perturbations go through a non-monotonous transient amplification. 
The realized trajectory of the transient very much depends on the projection of the perturbations onto the pseudo-eigenvectors \cite{Embree2005,Biancalani2017,Nicolaou2020}. 
Not all large Kreiss constant values should thus lead to a bubble,  as the market dynamics is more complex and cannot just be reduced to one source of influence. 
The mapping between large Kreiss constant and bubbles becomes rigorous when considered in terms of ensembles of price trajectories.
To show this, and to isolate the effect of discussions on Reddit, we insert the empirical Blackberry discussion forum influence network $\A(t)$ as input to our ABM. 
We simulate the resulting price time-series $100$ times and keep track of the average price as well as its average standard deviation (see Supplementary Note 6 for details). 
In the bottom plot of Figure \ref{fig:empirical}, one can observe indeed that price spikes coincide - in their ensemble average -  with peaks of the Kreiss constant, 
supporting our hypothesis that the non-normality in the Reddit discussion forum contributes to explain the observed price bubbles.
While meme stocks happen to be a prime example of price dynamics that are driven by opinion leaders \cite{Lucchini2022}, 
human interactions in general and financial markets in particular are known to be subject to non-normal interactions \cite{Fiske1992,Johnson2020,Chiarella2009}.
We thus hypothesize that the mechanism described in this article is an important contribution to the manifold observations of bubbles in socio-economic systems \cite{Lux1995,GislerSor11,sornette2014financial}. 

Our simulations exemplify the possibility to diagnose regimes of financial instabilities by measuring the evolution of the Kreiss constant of the underlying network of social interactions between traders.
Periods in which the Kreiss constant is large should be interpreted as prone to bubble regimes and large price volatility.
The mapping of the detection of (financial) instabilities to the measurement of the Kreiss constant improved 
conceptually and operationally on the previous approaches attempting to anticipate critical phase transitions \cite{Scheffer2009,Scheffer2012,Van2014,Ma2019,Lera2020},
which do not incorporate the ubiquitous non-normality of complex system dynamics.

\section*{Conclusions}
\label{sec:conclusions}

In most of the literature, financial and socio-economic bubbles have been thought of as being associated with special regimes where self-reinforcing interactions strengthen transiently towards a critical point and lead to some form of collective exuberance.
This has been formalized by models in physics, ecology and mathematics assuming the presence of an underlying phase transition, criticality, bifurcation or catastrophe. 
The focus put mainly on critical points is particularly surprising insofar as the existence of non-critical transients via non-normal dynamics has been well-established in other fields of science such as in hydrodynamic turbulence. 

Here, we have demonstrated, via agent-based model simulations and empirically,
that such transient phases of exuberance are also generic in real social systems ubiquitously characterized by non-normal properties of asymmetric hierarchical interactions and Ising-like imitative interactions.
Our key insight  is that we find seemingly critical behavior in the sub-critical regime as a result of transient collective behavior.
This is reminiscent of the behavior that would result in the agent-based model from the Ising-like interactions between noise traders being at or close to criticality in a normal network.
In such normal networks of interacting noise traders, if the Ising-like component is not crossing criticality, one does not observe bubbles.
It is the non-monotonic burst response to perturbation in the sub-critical regime of the non-normal network of interactive noise traders that create transient bubbles.

An important corollary is that financial bubbles should be expected as intrinsic, rather than abnormalities appearing in very special conditions.
This may explain their ubiquity in financial markets, from past centuries to the present.
Our results have been derived based on first-order principles that are not specific to our implementation but instead rely on the fact that, on short time-scales, opinion-leaders can significantly lead opinions astray from rational ones.
With these general assumptions and the broad applicability of models involving hierarchical and stochastic imitative interactions, 
our framework is expected to explain ubiquitous crowd-forming patterns and collective structures in general hierarchical social networks.

\section*{Methods}

\subsection*{Dynamics of Noise Traders}

In any ABM of financial markets, it is crucial to have at least two classes of traders. 
When market participants become too homogeneous, liquidity rarifies and no trades occur. 
We initiate the model with equal weight between the fundamentalists and the noise traders,  where their weights is determined by their financial capital. 
As their wealth evolves with time, the relative importance of each group varies with time. 
The specific description of the investment strategy of fundamentalists is found in the Supplementary Note 1 and refs. \cite{kaizoji2015super,westphal2020market}.

We label the set of $N$ noise traders (or nodes in the network) from $1, \ldots, N$. 
We denote the state of each node $i = 1, \ldots, N$ as $s^{i}=\pm{1}$. 
The transition probability $\pi$ that trader $i$ flips its state $s^i_t$ at time $t$ depends on the opinion of its in-neighbours, according to 
\begin{equation}
	\label{eq:trans_probability}
    \pi({s^i_{t+1}=-s^i_t})
    =
    \frac{p^{\pm}}{2}
    \left( 
    1-\kappa\frac{1}{k^{in}_{i}}{s}^{i}_{t}\sum_{j}a_{ij}{s^{j}_{t}}
    \right)
\end{equation}
where $p^{\pm}$ controls the average holding time (net of social influence) of each asset 
and the social coupling strength $\kappa$ determines the noise traders' susceptibility to social imitation.
Here, $\A= \{ a_{ij} \}$, where $a_{ij}=1$ if there exists a directed edge (influence) from node $j$ to node $i$, and $a_{ij}=0$ otherwise. 
The in-degree of node $i$ is the number of directed edges pointing to node $i$, which is given by $k_i^{in}=\sum_{j=1}^{N}a_{ij}$. 
In the Ising model, if node $i$ switches its state from time step $t$ to time step $t+1$, i.e. $s_{t+1}^i=-s_{t}^i$, the change of the value of node $i$'s state is $-2s^i_t$. 
Given the probability of node $i$ to switch its state according to \eqref{eq:trans_probability}, 
the average rate of change of the spin starting  in the state $s_{t}^i$ is given by $\Delta s^i_t = -2 ~ s_t^i ~\pi$.
We introduce the $N$-dimensional state column vector $\vec{s}(t)=  \left( s^1_t,s^2_t, ..., s^N_t \right)$.
Together with \eqref{eq:trans_probability}, the average rate of state transition can then be written as 
$
    \triangle{\vec{s}(t)}
    =
    \vec{s}(t+1)-\vec{s}(t) 
    = 
    p^{\pm}(\kappa\mathbf{\Lambda}\A-\mathbf{I})\vec{s}(t)
$
where $\mathbf{\Lambda}$ is an $N\times{N}$ diagonal matrix with $1 \left/ k^{in}_i \right.$ on the $i$-th diagonal entry
and $\mathbf{I}$ is the identity matrix. 
Introducing the matrix 
\begin{equation}
    \label{eq:M_matrix}
    \M \equiv p^{\pm}(\kappa\mathbf{\Lambda} \A-\mathbf{I})~,
\end{equation}
recovers the simple dynamical equation \eqref{eq:vectorial_state_function}.

\subsection*{A Primer on Non-Normality}
\label{sec:primer_non_normality}

Following a classic textbook \cite{Embree2005}, we briefly summarize the basic theory behind non-normal matrices (see Supplementary Note 2 for details).

Let $\M$ be an $(N \times N)$-matrix.
The set of all eigenvalues of $\M$ is called the spectrum $\sigma(\M)$.
A matrix is called \textit{normal} if $\M^T \M =  \M  \M^T$, 
and the spectral theorem asserts that each normal $\M$ has a set of $n$ pairwise orthonormal eigenvectors of $\M$. 
By contrast, if $\M$ is \textit{non-normal}, $\M^T \M \neq  \M  \M^T$,  {there exist some eigenvectors which are non-orthogonal to each other.}
Since symmetric matrices are always normal, it is a necessary, but not a sufficient condition that matrix \eqref{eq:M_matrix} contains directed interactions to be considered non-normal.

If $\lambda$ is an eigenvalue of $\M$,  the resolvent matrix $\M - \lambda \FM{I}$ is not invertible since there exists an eigenvector $\vec{v}$ with $(\M - \lambda \FM{I}) \vec{v} = 0$.
An alternative definition of the spectrum $\sigma(\M)$ is thus the set of points $\lambda \in \mathbb{C}$ where the resolvent matrix does not exit.            
But the question ``Does $\left( \M-\lambda \FM{I}  \right)^{-1}$ exist?'' is binary and may change from ``yes'' to ``no'' by just a tiny $\epsilon$-perturbation of $\lambda$. 
In the presence of noise, a better question to ask is whether $\left| \left| \left( \M-\lambda \FM{I}  \right)^{-1} \right| \right|$ is large with respect to some matrix norm $\nrm{ \cdot }$.
This leads to the definition of the $\epsilon$-\textit{pseudospectrum}, defined as the set of points where $\left| \left| \left( \M-\lambda \FM{I}  \right)^{-1} \right| \right|$ is large (larger then $\epsilon^{-1}$), 
or formally, 
$
\sigma_\epsilon(\M) \equiv \left\{ \lambda \in \mathbb{C} ~: ~\nrm{ \left( \M - \lambda \FM{I} \right)^{-1} } > \epsilon^{-1} \right\}. 
$
The $\epsilon$-pseudospectrum is the open subset of the complex plane bounded by the $\epsilon^{-1}$ level-curve of the norm of the resolvent. 
Intuitively, one can then assume that the $\epsilon$-pseudospectrum is closely confined around the eigenvalues of $\M$.
For normal matrices, this assumption is correct.
However, for non-normal matrices it is not, and $\left| \left| \left( \M - \lambda \FM{I} \right)^{-1} \right| \right|$ may be large even when $\lambda$ is far away from the spectrum
(Figure \ref{fig:non_normal_theory} (a) and (b)).

Consider the proportional growth equation
$
\left. \dd \vec{s} \right/ \dd t = \M \vec{s}
$
with explicit solution $\vec{s}(t) = e^{t \M} \vec{s}(0)$. 
It is well-known that the asymptotic behavior for $t \to \infty$ is governed by the largest real-part of all eigenvalues of $\M$. 
For the short-term behavior, $t \downarrow 0$,  it can be shown that
$
\td{~}{t} \nrm{ e^{t \M} }_{t=0} = \omega(\M) \equiv \text{sup}~ \sigma \left( \frac{1}{2} \left( \M + \M^T \right) \right)
$
where $\omega(\M)$ is called the \textit{numerical abscissa} of $\M$ (Figure \ref{fig:non_normal_theory}(c)).
Our main interest are, however, intermediate values of $t$. 
To describe such transient behavior, one has to consider the $\epsilon$-\textit{spectral abscissa} of a matrix $\M$ defined by 
$
\alpha_\epsilon(\M) = \sup \text{Re} \left( \sigma_\epsilon(\M) \right), 
$
i.e. the supremum of the real part of the $\epsilon$-pseudo-spectrum. 
An important special case is the spectral abscissa $\alpha(\M) \equiv \alpha_{\epsilon=0}(\M)$, defined as the largest real-part of all eigenvalues of $\M$.

We now consider the case where $\alpha(\M) < 0$, i.e. where the long-term behavior is asymptotically stable  (Figure \ref{fig:non_normal_theory}(c)),
but $\alpha_\epsilon(\M) > 0$ for some $\epsilon > 0$. 
In that case, the pseudospectra of $\M$ protrude significantly into the right-half plane of $\mathbb{C}$, such that the real-parts of the pseudo-spectrum remain positive (Figure \ref{fig:non_normal_theory} (e)). 
For any such non-normal matrix $\M$, the \textit{Kreiss constant}
\begin{equation}
	\mathcal{K}(\M) \equiv \sup_{\epsilon > 0} \frac{ \alpha_\epsilon(\M) }{ \epsilon }
	\label{eq:Kreiss_constant}
\end{equation}
is well-defined, and it can be shown that, for intermediate times (Figure \ref{fig:non_normal_theory}(c)), there is transient growth according to \eqref{eq:Kreiss_inequality}.

In contrast to the `single-shot' dynamics in which the system relaxes back to the stable state after a short-lived shock,
our model \eqref{eq:vectorial_state_function} is stochastic such that the trajectory may get perturbed before it can equilibrate. 
It has been shown that, in such stochastic models, non-normality will increase the variance of the fluctuations, producing amplifications beyond what is expected from its normal counterpart \cite{Farrell1994,Biancalani2017,Nicoletti2018,Nicoletti2019}.

Finally, we note that an alternative measure to characterize an operator's degree of non-normality is \textit{Henrici's departure from normality} 
\begin{equation}
	d_F(\A) =  \left. 
			\sqrt{ \nrm{ \A }_F^2 - \sum_{\lambda \in \sigma(\A)} \left| \lambda \right|^2 }
			\right/
			\nrm{ \A }_F^2.
	\label{eq:Henrici}
\end{equation}
Henrici's index is based on the observation that the Frobenius norm of a normal matrix is given by 
$\nrm{\A}_F^2 = \text{tr} \left( \A^T \A \right) = \sum_{\lambda \in \sigma(\A)} \left| \lambda \right|^2$.
It attains its minimum at zero once the matrix is normal and increases the more the matrix deviates from normality. 
For example, the Henrici's indices of the adjacency matrices $\A$ of the networks depicted in Figure 2 (a,b,c) are equal to 1, 0.9007 and 0.8057, respectively. 

As we show in the Supplementary Note 4 - at least within the realms of our application - the relationship between $\mathcal{K}$ and $d_F$ is strictly monotonous. 
Therefore all of our results presented below are also valid when expressed in terms of $d_F(\M)$.

\subsection*{Parametrization of Non-Normal Matrices}
\label{sec:primer_non_normality}

The non-normal network with a total of $N$ nodes is initialized with $N_0$ top nodes. 
Each newly added node receives $m$ in-edges, i.e. channels of communication through which it can be influenced. 
The source of each such edge is selected with probability proportional to the existing nodes' out-degree.
Once the $m$ source nodes of an added node are determined,  each of the $m$ newly formed directed edges may be reciprocated with some probability $\rho$. 
Based on empirical evidence from Reddit, we assume further that this probability $\rho$ is modulated by the hierarchical level $\ell$ of each node, $\rho = \rho(\ell)$.
We thus model the rate of reciprocity as a sigmoid-function
\begin{equation}
\rho(\ell) =  \frac{ \theta}{ 1 + e^{ -a(\ell(j)-b)}}  ~-~ \underbrace{ \frac{ \theta}{ 1 + e^{ -a(1-b)}} }_{\equiv \gamma}
\end{equation}
where $\theta$ now serves as an upper asymptotic bound. 
In the remainder of this article, we fix $\theta = 0.2110, a=2.552$ and $b=3.668$ 
as determined empirically below on meme-stock reddit data (Figure \ref{fig:example_network}(d)). 
The offset $\gamma \approx 0.0002$ has been added so that  nodes in the highest level of the hierarchy,  $\ell=1$,  are never reciprocated, $\rho(\ell=1)=0~ \forall \theta$. 
This offset is merely for convenience, so that we can keep the number of top nodes (`opinion leaders') fixed.
For large values of $\ell$, we converge to the constant rate of reciprocity $\theta - \gamma \approx \theta$ as in \cite{Asllani2018}. 
Parameter $\theta$ now has the interpretation of the asymptotic level of reciprocity at high levels $\ell$. 
A value of $\theta = 1$ implies that most, albeit not all edges are reciprocated.
Details are found in Supplementary Note 3.

Our algorithm to generate non-normal adjacency matrix $\A$ has six parameters: $N, N_0, m, \theta, a$ and $b$. 
The price dynamics from \eqref{eq:vectorial_state_function} is not directly governed by $\A$, but rather by the related matrix $\M$.
The two parameters $\kappa$ and $p^\pm$ allow us to control the characteristics of $\M$ for given $\A$. 
In the remainder of this article, we fix $p^{\pm} = 0.05$ and we tune $\kappa$.
We refer to the Supplementary Note 4 for a detailed list of parameters, including parameters for the fundamental agents and macro-economic quantities. 
These parameters serve to fix the overall price scale, but do not qualitatively affect our results.
This leaves us with a three-parameter model, $N_0, \theta$ and $\kappa$. 
Importantly for what follows, the Kreiss constant $\mathcal{K}$ is strictly decreasing in $\theta$ and increasing in $\kappa$, irrespective of $N_0$, as long as $N_0 \ll N$. 
Throughout this article, we constrain the parameter such that $\alpha_0(\M) < 0$, i.e. the asymptotic system dynamics is stable.

{
\balance
\tiny
\bibliographystyle{unsrt}
\bibliography{bibliography.bib}
}

\end{document}